\documentclass[12pt]{article}
\usepackage{amssymb,amsmath,epsfig}
\allowdisplaybreaks
\begin{document}

\title{\bf Analysis of Generalized Ghost Pilgrim Dark Energy in Non-flat FRW Universe}
\author{Abdul Jawad
\thanks{jawadab181@yahoo.com; abduljawad@ciitlahore.edu.pk}\\
Department of Mathematics, COMSATS Institute of\\ Information
Technology, Lahore, Pakistan.}

\date{}

\maketitle
\begin{abstract}
This work is based on pilgrim dark energy conjecture which states
that phantom-like dark energy possesses the enough resistive force
to preclude the formation of black hole. The non-flat geometry is
considered which contains the interacting generalized ghost pilgrim
dark energy with cold dark matter. Some well-known cosmological
parameters (evolution parameter ($\omega_{\Lambda}$) and squared
speed of sound) and planes ($\omega_{\Lambda}$-$\omega_{\Lambda}'$
and statefinder) are constructed in this scenario. The discussion of
these parameters is totally done through pilgrim dark energy
parameter ($u$) and interacting parameter ($d^2$). It is interesting
to mention here that the analysis of evolution parameter supports
the conjecture of pilgrim dark energy. Also, this model remains
stable against small perturbation in most of the cases of $u$ and
$d^2$. Further, the cosmological planes correspond to $\Lambda$CDM
limit as well as different well-known dark energy models.
\end{abstract}

\section{Introduction}

The accelerated expansion of the universe is one of the active topic
in cosmology since its prediction \cite{1}. It is suggested through
different cosmological and astrological data arisen from well-known
observational schemes \cite{2}-\cite{6} that this rapid expansion is
due to an unknown force termed as dark energy (DE). Despite of many
efforts from different observational and theoretical ways, the
problem of DE is still not well settled due to its unknown nature.
In order to justify the source of accelerating expansion (i.e., the
nature of DE) of the universe, two different approaches have been
adopted. One way is to modify the geometric part of Einstein-Hilbert
action (termed as modified theories of gravity) for the discussion
of expansion phenomenon \cite{7}. The second approach is to propose
the different forms of DE called dynamical DE models.

Upto now, different dynamical DE models have been proposed in two
different contexts such as quantum gravity and general relativity.
Holographic DE (HDE) model has been proposed in the framework of
quantum gravity on the basis of holographic principle \cite{8}. The
density of HDE model has the following form \cite{9}
\begin{equation*}
\rho_{\Lambda}=3m^{2}m^2_pL^{-2},
\end{equation*}
where $m$ is a specific constant, $m_{p}=(8\pi G)^{-\frac{1}{2}}$
termed as reduced Planck mass and $L$ represent the infrared (IR)
cutoff described the size of the universe. This density has been
derived on the basis of idea of Cohen et al. \cite{10} limit which
is stated as the vacuum energy (or the quantum zero-point energy) of
a system with size $L$ should always remain less than the mass of a
black hole (BH) with the same size due to the formation of BH in
quantum field theory. This idea is reconsidered by Wei \cite{11}
with the proposal of pilgrim dark energy (PDE).

According to Wei, the formation of BH can be avoided through
appropriate resistive force which is capable to prevent the matter
collapse. In this phenomenon, phantom-like DE can play important
role which possesses strong repulsive force as compare to
quintessence DE. The effective role of phantom-like DE onto the mass
of the BH in the universe has also been observed in many different
ways. The accretion phenomenon is one of them which favor the
possibility of avoidance of BH formation due to presence of
phantom-like DE in the universe. It has been suggested that
accretion of phantom DE (which is attained through family of
chaplygin gas models \cite{12}) reduces the mass of BH. On the other
hand, there also exists a possibility of increasing of BH mass due
to phantom energy accretion process which leads to the violation of
cosmic censorship hypothesis \cite{13}. Hence, this phenomenon is
still unresolved.

It is strongly believed that the presence of phantom DE in the
universe will force it towards big rip singularity. This represents
that the phantom-like universe possesses ability to prevent the BH
formation. The proposal of PDE model \cite{11} also works on this
phenomenon which states that phantom DE contains enough repulsive
force which can resist against the BH formation. Wei \cite{11}
developed cosmological parameters for PDE model with Hubble horizon
and provided different possibilities for avoiding the BH formation
through PDE parameter. He adopted different possible theoretical and
observational ways to make the BH free phantom universe. Also, PDE
via reconstruction scheme is discussed in modified theory of gravity
such as $f(T)$ gravity \cite{S}. The behavior of cosmological
parameters along with validity of generalized second law of
thermodynamics are explored as well.

In addition, we worked on PDE models interacting with cold dark
matter (CDM) and pointed different ways in order to meet the PDE
phenomenon \cite{14}-\cite{16}. In this work, the generalized ghost
version of PDE model so called GGPDE interacting with CDM is
considered in non-flat universe. In this context, different
cosmological parameters (EoS parameter and squared speed of sound)
and planes ($\omega_{\Lambda}$-$\omega_{\Lambda}'$ and statefinder)
are developed. The format of the paper is as follows. Section
\textbf{2} contains the basic cosmological scenario, whereas section
\textbf{3} explores above mentioned cosmological parameters and
planes. The concluding remarks of the results are given in the last
section.

\section{Non-flat FRW Universe and Basic Equations}

In this section, we provide the basic scenario of non-flat geometry
of the universe as well as interacting scenario of GGPDE and CDM.
The basic purpose of this work to visualize the effects of spatial
curvature on PDE conjecture. It is found that different
observational analysis favor the flat universe. However, there are
arguments through observational schemes about the presence of small
fraction of spatial fractional density in the total fractional
energy contents of the universe. In non-flat FRW universe, the first
Friedmann equation becomes
\begin{eqnarray}\label{1}
H^2+\frac{k}{a^{2}}&=&\frac{1}{3m^2_{pl}}(\rho_m+\rho_\Lambda),
\end{eqnarray}
where $\rho_m$ and $\rho_\Lambda$ appear as CDM and GGPDE densities.
Also, $k=-1,~0,~1 $ describe open, flat and closed universes,
respectively. In cosmological context, the total amount of energy
density is calculated in terms of fractional energy density. Thus,
Eq.(\ref{1}) can be written in terms of fractional form as
\begin{equation}\label{2}
1+\Omega_{k}=\Omega_m+\Omega_\Lambda,\quad
\Omega_{k}=\frac{k}{a^2H^2},\quad
\Omega_{m}=\frac{\rho_m}{3m^2_{pl}H^2},\quad
\Omega_{\Lambda}=\frac{\rho_\Lambda}{3m^2_{pl}H^2}.
\end{equation}

It is well-known that dynamical DE models play an important role in
describing the accelerated expansion of the universe. The Veneziano
ghost DE is one of the dynamical DE model which is defined as
follows \cite{17}
\begin{equation*}
\rho_\Lambda=\alpha H,
\end{equation*}
where $\alpha$ is a constant with dimension [energy]$^3$. This model
is proposed on the basis of Veneziano ghost of choromodynamics (QCD)
which helps in solving the $U(1)$ problem in QCD. The Veneziano
ghost (being unphysical in quantum field theory formulation in the
Minkowski spacetime) provides non-trivial physical effects in FRW
universe \cite{18}. Although, QCD ghost possesses small contribution
in describing vacuum energy density which is proportional to
$\Lambda^{3}_{QCD}H$ (here $\Lambda_{QCD}\sim100MeV$ is the smallest
QCD scale), but this contribution plays important role in the
discussion of evolutionary universe. It is also investigated that
this model also helps in alleviating two major problems of DE called
fine tuning and cosmic coincidence problem \cite{17,19}. Many
authors have investigated/tested this model through different
cosmological parameters theoretically \cite{20} and different
observational schemes \cite{21}.

It is observed that the Veneziano ghost field in QCD of the form
$H+O(H^2)$ has ability in producing enough vacuum energy to explain
the accelerated expansion of the universe \cite{22}, but only
leading term (i.e., $H$) involved in ordinary ghost DE model. It is
suggested \cite{23} that the contribution of the term $H^2$ in the
ordinary ghost DE may be useful in describing the early evolution of
the universe which is defined as follows
\begin{equation*}
\rho_{\Lambda}=\alpha H+\beta H^2,
\end{equation*}
here $\beta$ involves as a constant containing dimension
[energy]$^2$ and corresponding energy density is called generalized
ghost DE. Upto now, this model was investigated by different
cosmological parameters such as EoS parameter, deceleration,
$\omega_\Lambda-\omega'_\Lambda$, statefinder and squared speed of
sound etc. \cite{16,24,25}. Its generalized version in terms of PDE
is defined as follows \cite{16}
\begin{equation}\label{3}
\rho_{\Lambda}=(\alpha H+\beta H^2)^u,
\end{equation}
known as GGPDE.

We take interaction between GGPDE and CDM which follows the
equations of continuity as
\begin{eqnarray}\label{4}
\dot{\rho}_{m}+3H\rho_{m}=\Gamma,\quad
\dot{\rho}_{\Lambda}+3H(\rho_{\Lambda}+p_{\Lambda})=-\Gamma,
\end{eqnarray}
where $\Gamma$ is known as interaction term between CDM and GGPDE
possessing dynamical behavior. The unknown nature of DE as well as
CDM leads to the basic problem for the choice of interaction term.
It is difficult to describe interaction via first principle.
However, the continuity equation provides a clue about the form of
interaction, i.e., it must be a function of the product of energy
density and a term with units of time (such as Hubble parameter).
With this idea, different forms for interaction have been proposed.
We take the following form of this interaction term
\begin{eqnarray}\label{11B}
\Gamma=3d^2H\rho_{m},
\end{eqnarray}
with $d^2$ serves as interaction parameter which exchanges the
energy between CDM and DE components. This form of interaction term
has been explored for energy transfer through different cosmological
constraints. The sign of coupling constant decides the decay of
energies either DE decays into CDM (when the interacting parameter
is positive) or CDM decays into DE (when the interacting parameter
is negative). The present analysis from different aspects imply that
the phenomenon of DE decays into CDM which is more acceptable and
favors the observational data.

\section{Cosmological Parameters}

Here, we discuss the evolution of the Hubble parameter, the universe
and stability of the interacting model GGPDE. For this purpose, we
extract EoS parameter and squared speed of sound.

\subsection{Hubble Parameter}

By using Eqs.(\ref{1})-(\ref{11B}), we obtain the differential
equation in term of Hubble parameter as follows
\begin{eqnarray}\nonumber
\dot{H}(a)&=&\Omega_{m0}H_{0}^2(-3(1-d^2)H(a)a^{-3(1-d^2)})+(2H(a))a^{-2}
(2H(a)\\\label{H5}&-&\frac{u}{3}(\alpha H(a)+\beta
H(a)^2)^{u-1}(\alpha+\beta H(a)))^{-1}.
\end{eqnarray}
We solve it numerically for $H(a)$ and plot it against cosmic scale
factor $a$ for three different values of $u=0.5,~-0.5,~1$ as shown
in Figures \textbf{1-3}. We chose initial condition $H(1)=74$ and
other constant parameters are $d^2=0.02,~0.03,~0.04$ and
$\Omega_{m0}=0.27,~H_0= 74,~\alpha= -1.05,~\beta= 2.25$. It can be
observed through all plots for all values of interacting parameter
$d^2$ that $H(a)$ shows increasing behavior which is consistent with
the present day observations about the expanding of the universe.
Also, it can be observed that the trajectories of $H(a)$ remains in
the range $[74,75]$ which is consistent with the recent planck data
as obtained by Ade et al. \cite{R33}.
\begin{figure} \centering
\epsfig{file=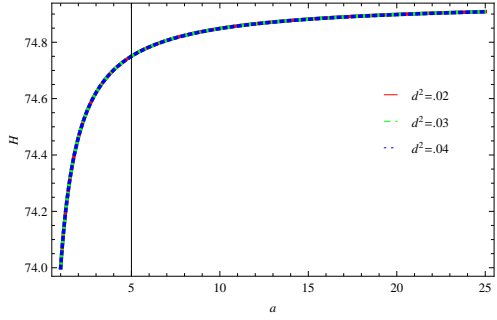,width=.50\linewidth}\caption{Plot of $H$
versus $a$ for GGPDE in non-flat universe with $u=0.5$.}
\end{figure}
\begin{figure} \centering
\epsfig{file=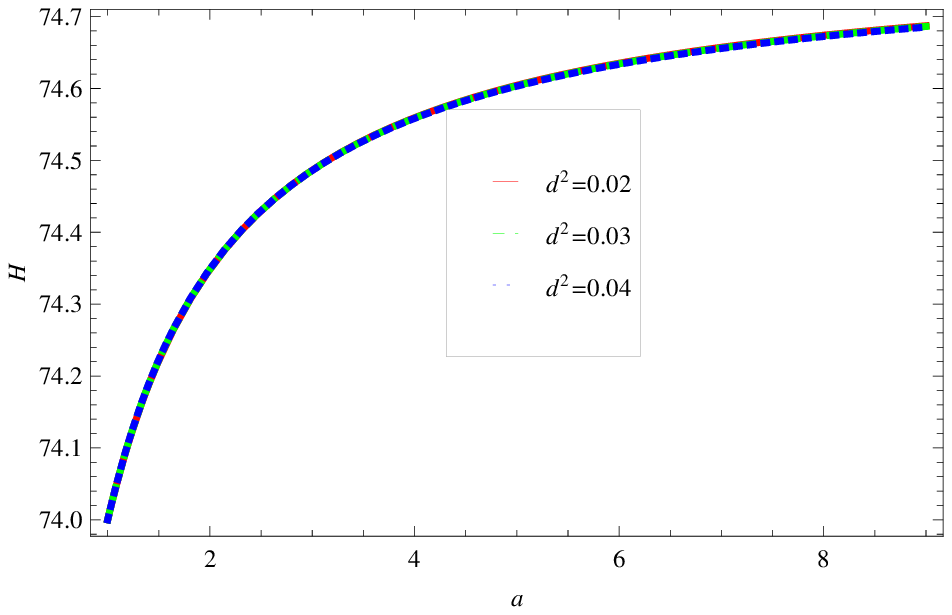,width=.50\linewidth}\caption{Plot of $H$
versus $a$ for GGPDE in non-flat universe with $u=-0.5$.}
\end{figure}
\begin{figure} \centering
\epsfig{file=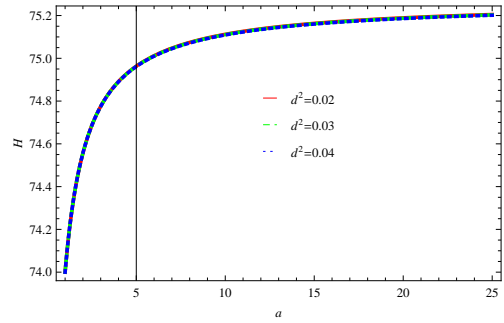,width=.50\linewidth}\caption{Plot of $H$ versus
$a$ in non-flat universe with $u=1$.}
\end{figure}

\subsection{The Equation of State Parameter}

In this scenario, EoS parameter takes the form
\begin{eqnarray}\nonumber
\omega_{\Lambda}&=&-1-d^2((\Omega_m)(\Omega_{\Lambda})^{-1}-(u(\alpha+
2\beta(H(a))))(3(H(a)))^{-1}\Omega_{m0}\\\nonumber
&\times&H_0^2(-3 (1-d^2)H(a)a^{-3
(1-d^2)})+(2(H(a)))a^{-2})(2(H(a))\\\label{5} &-&\frac{u}{3}(\alpha
H(a)+\beta H(a)^2)^{u-1}(\alpha+\beta H(a)))^{-1}.
\end{eqnarray}
We analyze the behavior of EoS parameter corresponding to three
different values of PDE parameter $u$, i.e., $u=0.5,~-0.5,~1$ and
keeping the same values of other constant parameters as shown in
Figures \textbf{4-6}. In order to observe the effects of interaction
parameter on PDE phenomenon, we take its three different values such
as $d^2=0.02,~0.03,~0.04$. In Figure \textbf{4} ($u=0.5$), it can be
observed that the EoS parameter starts from phantom region (with
comparatively large negative value) and goes towards lower negative
value of phantom region for all cases of interacting parameter. For
$u=-0.5$ (Figure \textbf{5}), it starts from quintessence phase  and
turns towards phantom region by crossing vacuum dominated era of the
universe for the cases ($d^2=0.02,~0.03$). However, it remains in
the phantom region for $d^2=0.04$. Also, Figure \textbf{6} provided
that EoS parameter starts comparatively high value of phantom region
and always remains in that region for all values of interacting
parameter. It can also be observed that EoS parameter attains high
phantom region with the increase of interacting parameter. The above
discussion shows that all the models provides fully support the PDE
phenomenon.
\begin{figure} \centering
\epsfig{file=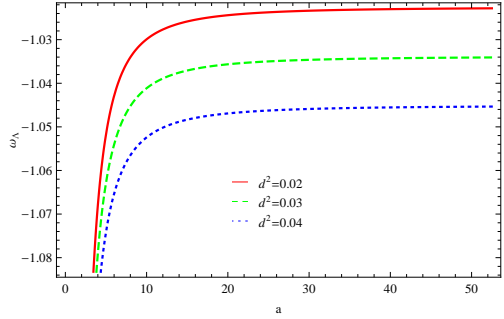,width=.50\linewidth}\caption{Plot of
$\omega_{\Lambda}$ versus $a$ for GGPDE in non-flat universe with
$u=0.5$.}
\end{figure}
\begin{figure} \centering
\epsfig{file=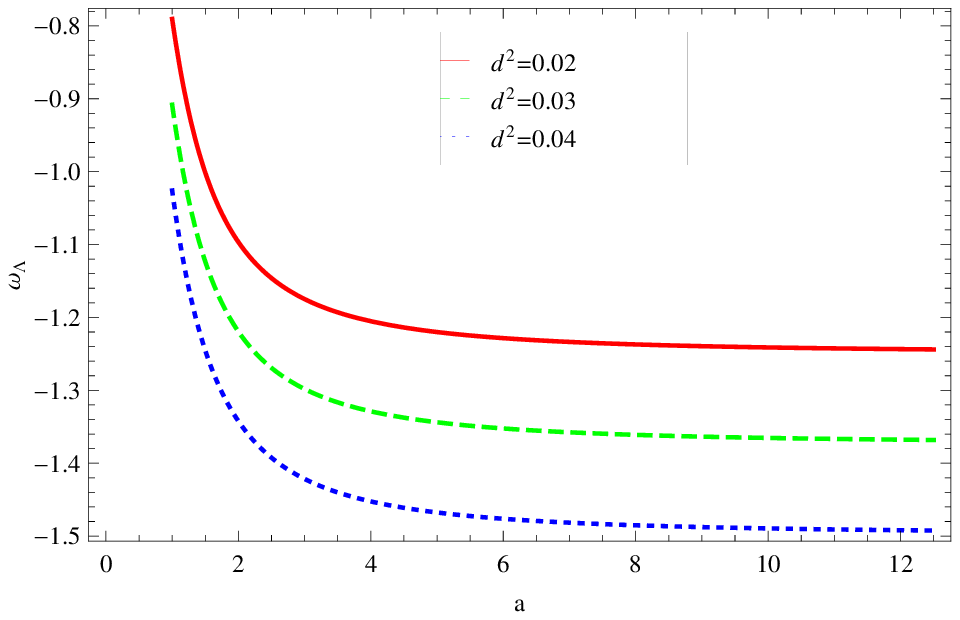,width=.50\linewidth}\caption{Plot of
$\omega_{\Lambda}$ versus $a$ for GGPDE in non-flat universe with
$u=-0.5$.}
\end{figure}
\begin{figure} \centering
\epsfig{file=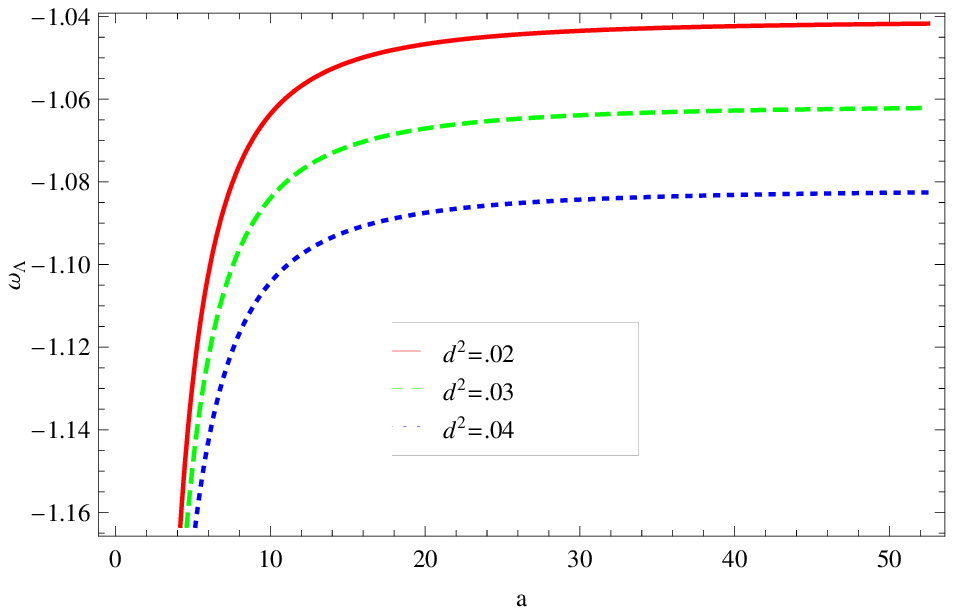,width=.50\linewidth}\caption{Plot of
$\omega_{\Lambda}$ versus $a$ in non-flat universe with $u=1$.}
\end{figure}

\subsection{Stability Analysis}

Now, we use squared speed of sound for the stability analysis of the
present interacting model. It is given by
\begin{equation}\label{6}
\upsilon_{s}^2=\frac{\dot{p}}{\dot{\rho}}=\frac{p'}{\rho'},
\end{equation}
By following \cite{16}, we obtain the following expression
\begin{eqnarray}\nonumber
\upsilon^2_{s}&=&-1+b-(3d^2(H(a)(\alpha+H(a)))^{-u}(1+a^2H(a)^2))a^{-2}\\\nonumber
&+&(u H(a)(\alpha+2\beta H(a))(2a+3a^{3d^2}(-1+d^2)H_0^2
\Omega_{m0}))\\\nonumber &\times&(a^3(-6H(a)^2+u(H(a)(\alpha +\beta
H(a)))^u))^{-1}+((2H(a)\\\nonumber &-&(u(H(a)(\alpha+\beta
H(a)))^u(3H(a))^{-1}) ((3ad^2H(a)^2(\alpha(-2\\\nonumber
&+&u)+2(-1+u)H(a))(2a+3a^{3
d^2}(-1+d^2)H_0^2\Omega_{m0}))((\alpha\\\nonumber &+&
H(a))(6H(a)^2-u(H(a)(\alpha+\beta H(a)))^u))^{-1}+(6d^2\\\nonumber
&\times&H(a)^2 (H(a)(\alpha+H[a]))^{-u}
(-3+a^2(-3H(a)^2+(H(a)\\\nonumber
&\times&(\alpha+H(a)))^u))(2a+3a^(3d^2)(-1+d^2)
H_0^2\Omega_{m0}))(6aH(a)^2 \\\nonumber &-&au(H(a)(\alpha +\beta
H(a)))^u)^{-1}+(3d^2uH(a)^3(H(a)(\alpha+H(a)))^{-1-u}
\\\nonumber &\times&(\alpha+2H(a))(-3+a^2(-3H(a)^2+(H(a)(\alpha+H(a)))^u))\\\nonumber &\times&(2a+
3a^(3d^2)(-1+d^2)H_0^2\Omega_{m0}))(-6aH(a)^2 +au(H(a)\\\nonumber
&\times&(\alpha+\beta H(a)))^u)^{-1}-(6\beta u H(a)^4(2a+
3a^{3d^2}(-1+d^2)\\\nonumber &\times&H_0^2\Omega_{m0})^2)(-6aH(a)^2
+au(H(a)(\alpha+\beta H(a)))^u)^{-2}+(3u H(a)^3 \\\nonumber
&\times&(\alpha+2\beta
H(a))(2a+3a^{3d^2}(-1+d^2)H_0^2\Omega_{m0})^2)(-6 a H(a)^2
\\\nonumber &+&au(H(a)(\alpha+\beta H(a)))^u)^{-2}+(3u H(a)^3
(\alpha+2\beta H(a))\\\nonumber &\times&(-6H(a)^2(\alpha+\beta
H(a))+u(H(a)(\alpha +\beta H(a)))^u(\alpha(-1+u)\\\nonumber
&+&\beta(-1+2u)H(a)))
(2a+3a^(3d^2)(-1+d^2)H_0^2\Omega_{m0})^2)(a^2(\alpha\\\nonumber
&+&\beta H(a))(-6H(a)^2+u(H(a)(\alpha+\beta
H(a)))^u)^3)^{-1}+6ad^2(H(a)\\\nonumber
&\times&(\alpha+H(a)))^{-u}(H(a)+(3H(a)^2(2a+
3a^{3d^2}(-1+d^2)H_0^2\Omega_{m0}))\\\nonumber
&\times&(a^2(6H(a)^2-u (H(a)(\alpha+\beta H(a)))^u)))-(u
H(a)^2(\alpha+2\beta H(a))\\\nonumber &\times&(a^2u(H(a)(\alpha
+\beta H(a)))^u(4a-9a^{3d^2}(-1+d^2)^2 H_0^2\Omega_{m0})\\\nonumber
&+&3H(a)((2a+3a^(3d^2)(-1+d^2)H_0^2
\Omega_{m0})^2+2a^2H(a)(-4a\\\nonumber &+&9a^{3d^2}(-1+d^2)^2
H_0^2\Omega_{m0}))))(-6aH(a)^2+au(H(a)\\\nonumber
&\times&(\alpha+\beta H(a)))^u)^2))(a u H(a)(2a+
3a^{3d^2}(-1+d^2)H_0^2\Omega_{m0})).
\end{eqnarray}
\begin{figure} \centering
\epsfig{file=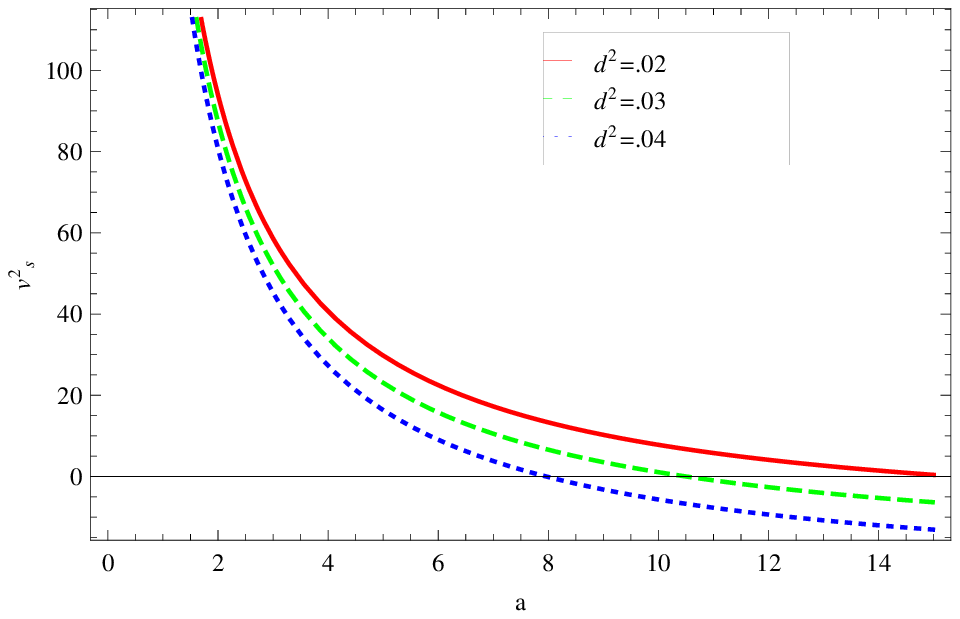,width=.50\linewidth}\caption{Plot of
$\upsilon^2_s$ versus $a$ for GGPDE in non-flat universe with
$u=0.5$.}
\end{figure}
\begin{figure} \centering
\epsfig{file=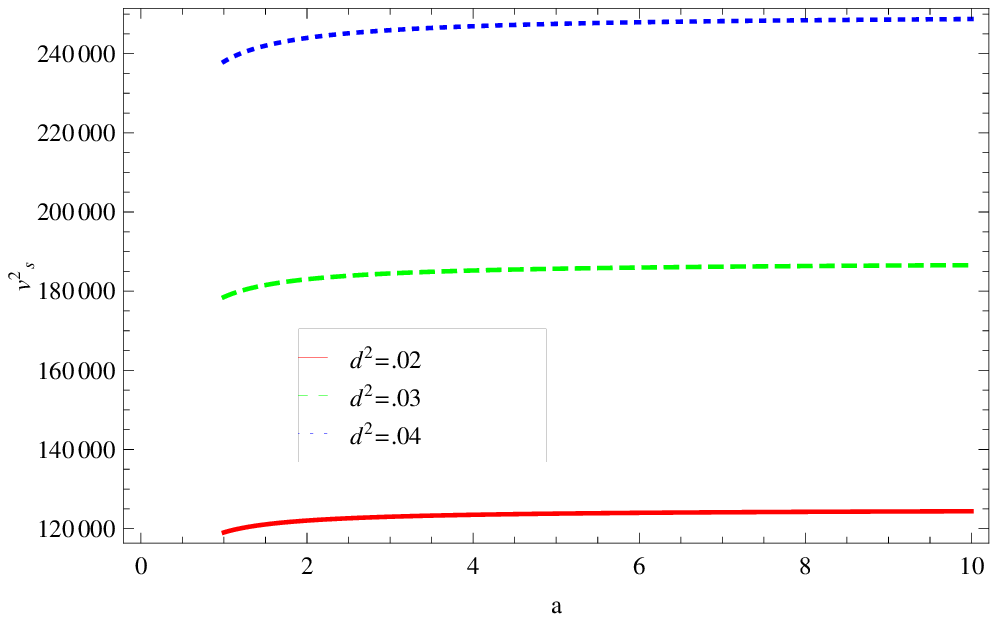,width=.50\linewidth}\caption{Plot of
$\upsilon^2_s$ versus $a$ for GGPDE in non-flat universe with
$u=-0.5$.}
\end{figure}
\begin{figure} \centering
\epsfig{file=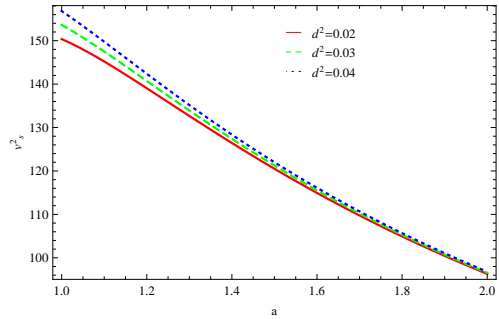,width=.50\linewidth}\caption{Plot of
$\upsilon^2_s$ versus $a$ in non-flat universe with $u=1$.}
\end{figure}

In order to analyze the behavior of squared speed of sound, we plot
the $\upsilon^2_{s}$ versus $a$ for its three different values,
i.e., $u=0.5,~-0.5,~1$ as shown in Figures \textbf{7-9}. In Figure
\textbf{7}, it is observed that GGPDE remains stable against small
perturbation at the present epoch as well as recent present epoch.
It can be viewed from Figure \textbf{8} ($u=-0.5$) that the GGPDE
model exhibits stability for all values of interacting parameter in
this scenario due to positive behavior of squared speed of sound. In
case of $u=1$ (Figure \textbf{9}), the squared speed of sound also
exhibits stability of the model for all cases of $d^2$.

\subsection{$\omega_{\Lambda}-\omega'_{\Lambda}$ Analysis}

The $\omega_{\Lambda}-\omega'_{\Lambda}$ plane is used to discuss
the dynamical property of DE models, where $\omega'_{\Lambda}$ is
the evolutionary form of $\omega_{\Lambda}$ (prime represents
derivative with respect to $\ln a$). Caldwell and Linder \cite{75}
firstly proposed this method for analyzing the behavior of
quintessence scalar field DE model. They pointed out that
$\omega_{\Lambda}-\omega'_{\Lambda}$ plane for quintessence model
with scalar field potential asymptotically approaching to zero, can
be divided into two categories of thawing and freezing regions. In
thawing region, EoS parameter begins nearly from $-1$ and increases
with time while its evolution remains positive. In freezing region,
EoS parameter remains negative and decreases with time while its
evolution also remains negative. In other words, the thawing region
is described as $\omega'_{\Lambda}>0$ for $\omega_{\Lambda}<0$ while
freezing region as $\omega'_{\Lambda}<0$ for $\omega_{\Lambda}<0$.
Later, this study was extended for examining the dynamical nature of
various DE models such as more general form of quintessence
\cite{76}, quintom \cite{77}, phantom \cite{78}, holographic
\cite{79}, polytropic DE \cite{80} and PDE \cite{14}-\cite{16}
models. Differentiating $\Omega_{k}$ and $\Omega_{\Lambda}$ with
respect to $x$ and after some manipulations, we get
\begin{eqnarray}\nonumber
\Omega'_k&=&-2\Omega_k(1+((\Omega_k-(1+\Omega_k-\Omega_\Lambda)+3d^2(\Omega_\Lambda-(1+\Omega_k
-\Omega_\Lambda)))(\alpha\\\label{2} &+&\beta H))(2(\alpha+\beta
H)-u\Omega_\Lambda(\alpha+2\beta H))^{-1}),\\\nonumber
\Omega'_k&=&\Omega_\Lambda(((\Omega_k-(1+\Omega_k-\Omega_\Lambda)+3d^2(\Omega_\Lambda-(1+\Omega_k
-\Omega_\Lambda)))(u(\alpha+\beta H)\\\label{2} &-&\alpha-\beta
H))(2(\alpha+\beta H)-u \Omega_\Lambda(\alpha+ 2\beta H))^{-1})
\end{eqnarray}
By taking the derivative of Eq.(\ref{5}) and using the above
expression, we get the evolutionary form of $\omega_{\Lambda}$ as
follows
\begin{eqnarray}\nonumber
\omega'_{\Lambda}&=&\frac{1}{a^5H(a)}((3ad^2H(a)^2(\alpha(-2+u)+2
(-1+u)H(a))(2a\\\nonumber&+&3a^{3d^2}(-1+d^2)H_0^2\Omega_{m0}))((\alpha+H(a))(6H(a)^2
-u(H(a)\\\nonumber&\times&(\alpha+\beta
H(a)))^u))^{-1}+(6d^2H(a)^2(H(a)(\alpha+H(a)))^{-u}(-3\\\nonumber&+&a^2(-3H(a)^2
+(H(a)(\alpha+H(a)))^u))(2a+3a^{3d^2}(-1+d^2)\\\nonumber&\times&H_0^2\Omega_{m0}))/(6aH(a)^2
-au(H(a)(\alpha+\beta
H(a)))^u)+(3d^2u\\\nonumber&\times&H(a)^3(H(a)(\alpha+H(a)))^{-1-u}(\alpha+2H(a))(-3+a^2\\\nonumber&\times&(-3H(a)^2
+(H(a)(\alpha+H(a)))^u))(2a+3a^{3d^2}(-1+d^2)\\\nonumber&\times&H_0^2\Omega_{m0}))(-6aH(a)^2+au
(H(a)(\alpha+\beta H(a)))^u)^{-1}-(6\beta\\\nonumber&\times& u
H(a)^4(2a+3a^{3d^2}(-1+d^2)H_0^2\Omega_{m0})^2)(-6 a
H(a)^2+au\\\nonumber&\times&(H(a)(\alpha+\beta
H(a)))^u)^{-2}+(3uH(a)^3(\alpha+2\beta
H(a))(2a\\\nonumber&+&3a^{3d^2}(-1+d^2)H_0^2\Omega_{m0})^2)(-6aH(a)^2+au(H(a)
(\alpha+\beta H(a)))^u)^{-2}\\\nonumber&+&(3uH(a)^3(\alpha+2\beta
H(a))(-6H(a)^2(\alpha+\beta H(a))+u(H(a)(\alpha\\\nonumber&+&\beta
H(a)))^u
(\alpha(-1+u)+\beta(-1+2u)H(a)))(2a+3a^{3d^2}(-1+d^2)\\\nonumber&\times&H_0^2
\Omega_{m0})^2)(a^2(\alpha+\beta H(a))(-6H(a)^2+u(H(a)(\alpha+ \beta
H(a)))^u)^3)^{-1}\\\nonumber&+&6ad^2(H(a)(\alpha+H(a)))^{-u}
(H(a)+(3H(a)^2(2a+3a^{3d^2}(-1+d^2)\\\nonumber&\times&H_0^2\Omega_{m0}))(a^2(6H(a)^2-u
(H(a)(\alpha+\beta H(a)))^u))^{-1})-(u
H(a)^2\\\nonumber&\times&(\alpha+2\beta H(a))(a^2u
(H(a)(\alpha+\beta
H(a)))^u(4a-9a^{3d^2}(-1+d^2)^2\\\nonumber&\times&H_0^2
\Omega_{m0})+3H(a)((2a+3a^{3d^2}(-1+d^2)H_0^2\Omega_{m0})^2+2a^2H(a)\\\nonumber&\times&
(-4a+9a^{3d^2}(-1+d^2)^2H_0^2\Omega_{m0}))))(-6aH(a)^2+au(H(a)\\\nonumber&\times&
(\alpha+\beta H(a)))^u)^{-2}).
\end{eqnarray}
\begin{figure} \centering
\epsfig{file=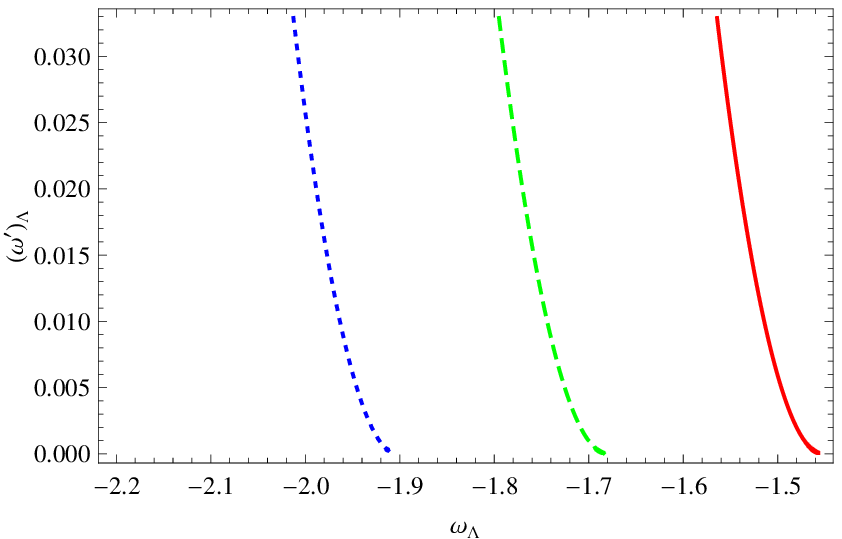,width=.50\linewidth}\caption{Plot of
$\omega_{\Lambda}-\omega_{\Lambda}'$ for GGPDE in non-flat universe
with $u=0.5$.}
\end{figure}
\begin{figure} \centering
\epsfig{file=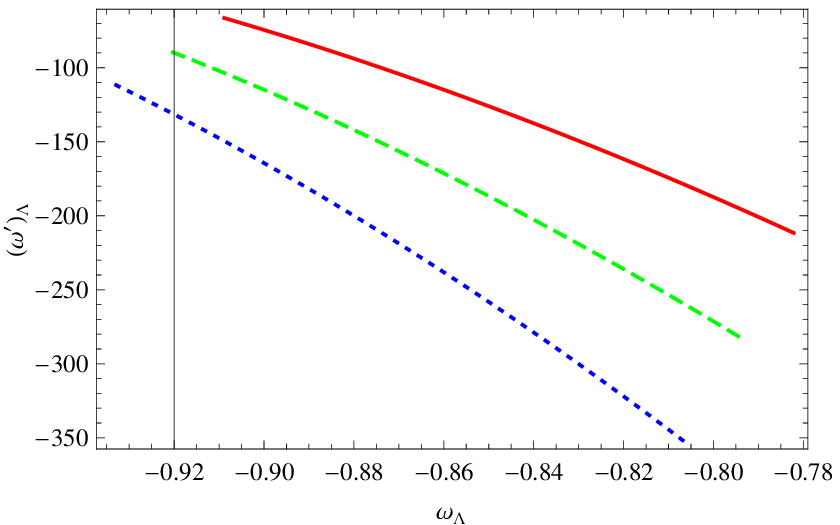,width=.50\linewidth}\caption{Plot of
$\omega_{\Lambda}-\omega_{\Lambda}'$ for GGPDE in non-flat universe
with $u=-0.5$.}
\end{figure}
\begin{figure} \centering
\epsfig{file=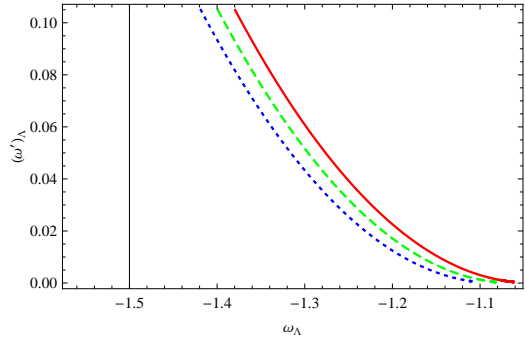,width=.50\linewidth}\caption{plot of
$\omega_{\Lambda}-\omega_{\Lambda}'$ in non-flat universe with
$u=1$.}
\end{figure}

The $\omega_{\Lambda}-\omega'_{\Lambda}$ plane for the current DE
model is constructed by plotting the $\omega'_{\Lambda}$ versus
$\omega_{\Lambda}$ for three different values of $u$ as shown in
Figures \textbf{10-12}. The specific values of other constant are
the same as above plots. Figures \textbf{10} and \textbf{12} provide
thawing region while Figure \textbf{11} exhibits freezing region.
The $\Lambda$CDM limit, i.e.,
$(\omega_{\Lambda},\omega'_{\Lambda})=(-1,0)$ only achieved for
$u=0.5$ with ($d^2=0$) as shown in Figure \textbf{10}. Hence,
$\omega_{\Lambda}-\omega'_{\Lambda}$ plane provides consistent
behavior with the present day observations in all cases of $u$.

\subsection{Statefinder Parameters}

The statefinder parameters depend upon two well-known basic
geometric parameters such as Hubble and deceleration which measure
expansion history of the universe. The deceleration parameter is
defined as
\begin{equation}\label{1.9.1}
q=-\frac{\ddot{a}}{aH^2}=-\left(1+\frac{\dot{H}}{H^2}\right).
\end{equation}
Notice that $\dot{a}>0$ represents the expansion of the universe
which yields $H>0$ while $\ddot{a}>0$ exhibits accelerated expansion
of the universe which provides negative deceleration parameter
($q<0$). Thus the negative value of deceleration parameter
demonstrates accelerated expansion of the universe, its positive
value shows decelerated phase of the universe while its zero value
shows uniform expansion of the universe.

A large number of DE models have been proposed for elaborating the
phenomenon of DE in the accelerated expansion of the universe. It is
necessary to differentiate these models so that one can decide which
one provides better explanation for the current status of the
universe. Since various DE models exhibit the same present value of
the deceleration and Hubble parameter, so these parameters could not
be able to discriminate the DE models. For this purpose, Sahni et
al. \cite{80} introduced two new dimensionless parameters by
combining the Hubble and deceleration parameters which are expressed
as
\begin{eqnarray}\label{1.9.2}
r=\frac{\dddot{a}}{aH^3},\quad s=\frac{r-1}{3(q-\frac{1}{2})}.
\end{eqnarray}

These parameters have geometrical diagnostic due to their total
dependence on the expansion factor. The statefinders are useful in
the sense that we can find the distance of a given DE model from
$\Lambda$CDM limit. The well-known regions described by these
cosmological parameters are as follows: $(r,s)=(1,0)$ indicates
$\Lambda$CDM limit, $(r,s)=(1,1)$ shows CDM limit, while $s>0$ and
$r<1$ represent the region of phantom and quintessence DE eras. For
non-flat universe, the above parameters turn out to be
\begin{eqnarray}\label{1.9.3}
r=\frac{\dddot{a}}{aH^3},\quad
s=\frac{r-\Omega_{tot}}{3(q-\frac{\Omega_{tot}}{2})}.
\end{eqnarray}
Moreover, $r$ can be expressed in terms of Hubble parameter as
\begin{equation}\label{1.9.4}
r=\frac{\ddot{H}}{H^3}-3q-2.
\end{equation}
With the help of Eqs.(\ref{1.9.1}) and (\ref{1.9.4}), one can write
\begin{equation}\label{1.9.5}
r=2q^{2}+q-\frac{\dot{q}}{H}.
\end{equation}

By following the procedure of \cite{15}, we can get statefinders as
\begin{eqnarray}\nonumber
r&=&1+(a^2H(a)^2)^{-1}+3(2H(a)^2)^{-1}(\alpha
H(a)+H(a)^2)^u(-1-3d^2H(a)^2\\\nonumber&\times& (\alpha
H(a)+H(a)^2)^{-u}(1+(a^2H(a)^2)^{-1}-(\alpha
H(a)+H(a)^2)^u\\\nonumber&\times&(3H(a)^2)^{-1})-(u(\alpha+2\beta
H(a))((2H(a))a^{-2}-3a^{-3(1-d^2)}
(1-d^2)\\\nonumber&\times&H_0^2H(a)\Omega_{m0}))(3H(a)(2H(a)-
3^{-1}u(\alpha+\beta H(a))(\alpha H(a)+\beta\\\nonumber&\times&
H(a)^2)^{-1+u}))^{-1})(-3d^2H(a)^2(\alpha
H(a)+H(a)^2)^{-u}(1+(a^2H(a)^2)^{-1}\\\nonumber&-&(\alpha
H(a)+H(a)^2)^u(3H(a)^2)^{-1})+d^2(1-3H(a)^2(\alpha
H(a)\\\nonumber&+&H(a)^2)^{-u}(1+(a^2H(a)^2)^{-1}-(\alpha
H(a)+H(a)^2)^u(3H(a)^2)^{-1}))\\\nonumber&-&(u(\alpha+2 \beta
H(a))((2H(a))a^{-2}-3a^{-3(1-d^2)}(1-d^2)H_0^2
H(a)\Omega_{m0}))\\\nonumber&\times&(
3H(a)(2H(a)-3^{-1}u(\alpha+\beta H(a))(\alpha H(a)+\beta
H(a)^2)^{-1+u}))^{-1})\\\nonumber&-&(2a^4H(a)^3)^{-1}(\alpha
H(a)+H(a)^2)^u((3ad^2H(a)^2(\alpha(-2
+u)+2(-1\\\nonumber&+&u)H(a))(2a+3a^{3d^2}(-1+d^2)H_0^2\Omega_{m0}))
((\alpha+H(a))(6H(a)^2 \\\nonumber&-&u(H(a)(\alpha+\beta
H(a)))^u))^{-1}
+(6d^2H(a)^2(H(a)(\alpha+H(a)))^{-u}\\\nonumber&\times&(-3+a^2(-3H(a)^2+(H(a)
(\alpha+H(a)))^u))(2a+3a^{3d^2}(-1+d^2)\\\nonumber&\times&H_0^2
\Omega_{m0}))(6aH(a)^2-au(H(a)(\alpha+\beta
H(a)))^u)^{-1}+(3d^2uH(a)^3
\\\nonumber&\times&(H(a)(\alpha+H(a)))^{-1-u}
(\alpha+2H(a))(-3+a^2(-3H(a)^2+(H(a)\\\nonumber&\times&(\alpha+H(a)))^u))(2a+3a^{3
d^2}(-1+d^2)H_0^2\Omega_{m0}))(-6aH(a)^2+
au\\\nonumber&\times&(H(a)(\alpha+\beta H(a)))^u)-(6\beta u
H(a)^4(2a+ 3a^(3d^2)(-1+d^2)H_0^2
\\\nonumber&\times&\Omega_{m0})^2)(-6aH(a)^2 +au(H(a)(\alpha+\beta
H(a)))^u)^{-2}+(3uH(a)^3\\\nonumber&\times&(\alpha +2\beta
H(a))(2a+3a^{3d^2}(-1+d^2)H_0^2\Omega_{m0})^2)(-6 a
H(a)^2\\\nonumber&+&au(H(a)(\alpha+\beta
H(a)))^u)^{-2}+(3uH(a)^3(\alpha+2\beta
H(a))(-6H(a)^2\\\nonumber&\times&(\alpha+\beta
H(a))+u(H(a)(\alpha+\beta H(a)))^u
(\alpha(-1+u)+\beta(-1+2u)\\\nonumber&\times&H(a)))(2a+3a^{3d^2}(-1+d^2)
H_0^2\Omega_{m0})^2)(a^2(\alpha+\beta
H(a))(-6H(a)^2\\\nonumber&+&u(H(a)(\alpha+\beta H(a)))^u)^3)^{-1}+
6ad^2(H(a)(\alpha+H(a)))^{-u}(H(a)\\\nonumber&+&(3
H(a)^2(2a+3a^{3d^2}(-1+d^2)H_0^2\Omega_{m0}))(a^2
(6H(a)^2-u(H(a)(\alpha\\\nonumber&+&\beta H(a)))^u))^{-1})-(u
H(a)^2(\alpha +2\beta H(a))(a^2u(H(a)(\alpha+\beta
H(a)))^u\\\nonumber&\times&(4a-9a^(3d^2)(-1+
d^2)^2H_0^2\Omega_{m0})+3H(a)((2a+3a^{3d^2}
(-1+d^2)\\\nonumber&\times&H_0^2\Omega_{m0})^2+2a^2H(a)(-4a+9a^{3d^2}(-1+d^2)^2H_0^2\Omega_{m0}))))(-6aH(a)^2
\\\nonumber&+&au(H(a)(\alpha+\beta H(a)))^u)^{-2}),
\\\nonumber s&=&-3d^2H(a)^2(\alpha
H(a)+H(a)^2)^{-u}(1+1(a^2H(a)^2)^{-1}-(\alpha
H(a)+H(a)^2)^u\\\nonumber&\times&(3H(a)^2))+d^2(1-3H(a)^2(\alpha
H(a)+H(a)^2)^{-u}(1+1(a^2H(a)^2)^{-1}\\\nonumber&-&(\alpha
H(a)+H(a)^2)^u(3H(a)^2)^{-1}))-(u(\alpha+2\beta
H(a))((2H(a))a^{-2}\\\nonumber&-&3a^{-3(1-d^2)}(1-d^2)H_0^2H(a)\Omega_{m0}))(3H(a)(2H(a)-3^{-1}
u(\alpha+\beta H(a))\\\nonumber&\times&(\alpha H(a)+\beta
H(a)^2)^(-1+u)))^{-1}-((3ad^2
H(a)^2(\alpha(-2+u)+2\\\nonumber&\times&(-1+u)H(a))(2a+3a^(3d^2)(-1+d^2)H_0^2\Omega_{m0}))((\alpha+H(a))
(6H(a)^2\\\nonumber&-&u(H(a)(\alpha+\beta
H(a)))^u))^{-1}+(6d^2H(a)^2(H(a)(\alpha+
H(a)))^{-u}(-3\\\nonumber&+&a^2(-3H(a)^2+(H(a)(\alpha+
H(a)))^u))(2a+3a^{3d^2}
(-1+d^2)H_0^2\Omega_{m0}))\\\nonumber&\times&(6aH(a)^2-au(H(a)(\alpha+\beta
H(a)))^u)^{-1}+(3d^2uH(a)^3(H(a)\\\nonumber&\times&(\alpha+H(a)))^{-1-u}(\alpha+2H(a))(-3+
a^2(-3H(a)^2+(H(a)(\alpha\\\nonumber&+&H(a)))^u))(2a+3a^{3d^2}(-1+d^2)H_0^2
\Omega_{m0}))(-6 a H(a)^2+au(H(a)(\alpha\\\nonumber&+&\beta
H(a)))^u)^{-1}-(6\beta u H(a)^4(2a+3a^{3d^2}(-1+d^2) H_0^2
\Omega_{m0})^2)(-6aH(a)^2\\\nonumber&+&a u (H(a)(\alpha+\beta
H(a)))u)^{-2}+(3uH(a)^3(\alpha+2\beta
H(a))(2a+3a^{3d^2}(-1\\\nonumber&+&d^2)H_0^2\Omega_{m0})^2)(-6aH(a)^2+au(H(a)(\alpha
+\beta H(a)))^u)^{-2}+(3uH(a)^3 \\\nonumber&\times&(\alpha+2\beta
H(a))(-6H(a)^2 (\alpha+\beta H(a))+u(H(a)(\alpha+\beta
H(a)))^u(\alpha(-1\\\nonumber&+&u+
\beta(-1+2u)H(a)))(2a+3a^{3d^2}(-1+d^2)H_0^2\Omega_{m0})^2)(a^2(\alpha
+\beta\\\nonumber&\times& H(a))(-6H(a)^2+u(H(a)(\alpha+\beta
H(a)))^u)^3)^{-1}+6ad^2(H(a)(\alpha\\\nonumber&+&H(a)))^{-u}(H(a)+(3H(a)^2(2a+3
a^{3d^2}(-1+d^2)H_0^2\Omega_{m0}))(a^2(6H(a)^2\\\nonumber&-&u(H(a)(\alpha+\beta
H(a)))^u))^{-1})-(u H(a)^2(\alpha+2\beta
H(a))(a^2u(H(a)(\alpha\\\nonumber&+& \beta
H(a)))^u(4a-9a^{3d^2}(-1+d^2)^2H_0^2
\Omega_{m0})+3H(a)((2a+3a^{3d^2}(-1+d^2)\\\nonumber&\times&H_0^2
\Omega_{m0})^2+2a^2H(a)(-4a+9a^{3d^2}(-1+d^2)^2H_0^2
\Omega_{m0}))))(-6aH(a)^2\\\nonumber&+&au(H(a)(\alpha+\beta
H(a)))^u)^{-2})(3a^4 H(a)(-1-3d^2H(a)^2(\alpha
H(a)\\\nonumber&+&H(a)^2)^{-u}(1+1(a^2H[a]^2)^{-1}- (\alpha H(a)+
H(a)^2)^u(3H(a)^2)^{-1})-(u(\alpha\\\nonumber&+&2 \beta H(a))((2
H(a))a^{-2}-3a^{-3(1-d^2)}(1-d^2)H_0^2H(a)\Omega_{m0}))(3H(a)\\\nonumber&\times&(2H(a)
-3^{-1}u(\alpha+\beta H(a))(\alpha H(a)+\beta
H(a)^2)^{-1+u}))^{-1}))^{-1}.
\end{eqnarray}
\begin{figure} \centering
\epsfig{file=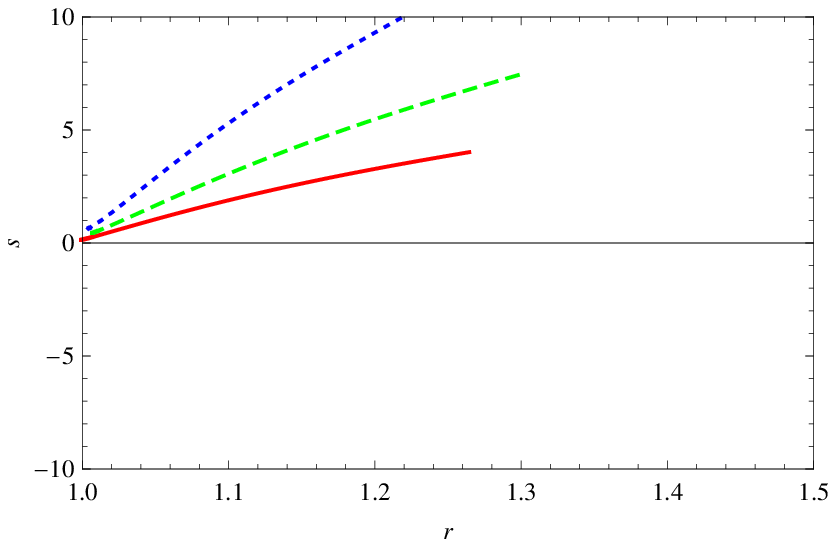,width=.50\linewidth}\caption{Plot of $r-s$
for GGPDE in non-flat universe with $u=0.5$.}
\end{figure}
\begin{figure} \centering
\epsfig{file=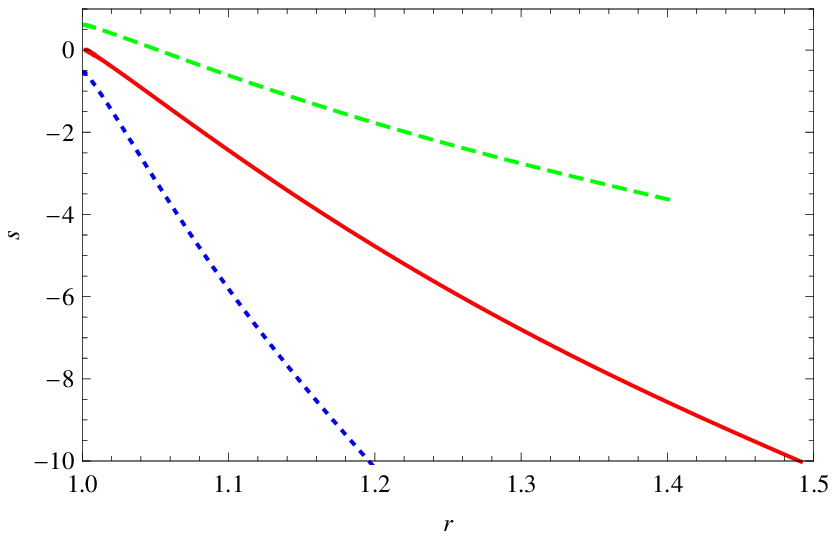,width=.50\linewidth}\caption{$r-s$ for
GGPDE in non-flat universe with $u=-0.5$.}
\end{figure}
\begin{figure} \centering
\epsfig{file=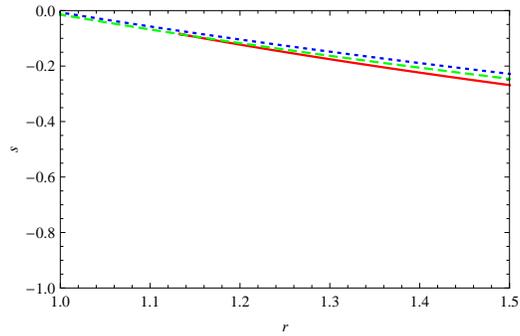,width=.50\linewidth}\caption{$r-s$ in non-flat
universe with $u=1$.}
\end{figure}
The $r-s$ plane corresponding to this scenario is shown in Figure
\textbf{13-15}. It is observed that the trajectories of $r-s$ plane
for all cases of interacting parameter corresponds to $\Lambda$CDM
model for $u=0.5$ as shown in Figure \textbf{13}. However, the
trajectories of $r-s$ meet $\Lambda$CDM limit for only $d^2=0$ (in
case of $u=-0.5$) and $d^2=0.02,~0.03$ (in case of $u=-0.5$) as
shown in Figures \textbf{14} and \textbf{15}, respectively. Also,
the trajectories coincide with the chaplygin gas model in all cases
of $u$.

\section{Results and Discussions}

It is well-known that total energy density of the universe contains
contribution of its different constituents in ratio of
$\Omega_k<\Omega_m<\Omega_\Lambda$ (the cosmic curvature density
$\Omega_k$ is found to be fractional). The early inflation era
indicates that the universe is non-flat if the number of e-folding
are small. It is predicted through many inflationary models that the
order of spatial curvature ($|\Omega_k|$) in the universe should be
less than $10^{-5}$ (but there are also exist some models which
allow larger curvature) \cite{38a,38b}. Also, the bound on EoS
parameter of different DE models was established in the non-flat
scenario of the universe by using observations of SNe Ia, BAO and
CMBR \cite{38c}. The range $(-0.2851,0.0099)$ of $\Omega_k$ at
$95\%$ confidence level was obtained with the help of WMAP five year
data \cite{38b} which was improved upto $(-0.0181,0.0071)$ by using
the data of BAO and SNe Ia. The range $-0.0133<\Omega_k<0.0084$ was
obtained by using latest WMAP $7$-years \cite{38d}.

An independent analysis of non-flat models based on time-delay
measurements of two strong gravitational lens systems, combined with
seven-year WMAP data, give consistent and nearly competitive
constraints of $\Omega_k=0.003^{+0.005}_{-0.006}$ \cite{38e}.
Recently, Ade et al. \cite{R33} (Planck data) found following
constraints on $\Omega_k$
\begin{eqnarray*}
100\Omega_{k}&=&-4.2^{+4.3}_{-4.8}~~~~~~~\text{(Planck+WP+highL)},\\
100\Omega_{k}&=&-0.10^{+4.8}_{-0.65},~~~\text{(Planck+lensing+WP
+highL)}
\end{eqnarray*}
These constraints are improved substantially by the addition of BAO
data which are
\begin{eqnarray*}
100\Omega_{k}&=&-0.05^{+0.65}_{-0.66}~~~~~~~\text{(Planck+WP+highL+BAO)},\\
100\Omega_{k}&=&-1.0^{+1.8}_{-1.9},~~~\text{(Planck+lensing+WP
+highL+BAO)}
\end{eqnarray*}
These limits are consistent with (and slightly tighter than) the
results reported by Hinshaw et al. \cite{Hin} from combining the
nine-year WMAP data with high resolution CMB measurements and BAO
data. Also, details about the curvature in the universe is given in
the recent Planck data \cite{R33}.

The above discussion motivate us to explore PDE phenomenon with
generalized ghost DE model in non-flat FRW universe. For this
purpose, two versatile cosmological parameters have been extracted
such as EoS parameter and squared speed of sound for analyzing the
behavior of evolution of the universe and stability of the model.
Also, two cosmological planes have been constructed for providing
the comparison of this model with other well-known DE models. The
discussion of the developed parameters are summarized as follows:
\begin{itemize}

\item Two recent analysis have greatly improved the precision of
the cosmic distance scale. Riess et al. \cite{Riess} use HST
observations of Cepheid variables in the host galaxies of eight SNe
Ia to calibrate the supernova magnitude-redshift relation. Their
"best estimate" of the Hubble constant, from fitting the calibrated
SNe magnitude-redshift relation, is
\begin{eqnarray*}
H_0&=&73.8\pm2.4km s^{-1}Mpc^{-1}~~~~~~~\text{(Cepheids+SNe Ia)},\\
\end{eqnarray*}
where the error is $1\sigma$ level and includes known sources of
systematic errors. Freedman et al. \cite{Freedman}, as part of the
Carnegie Hubble Program, use Spitzer Space Telescope mid-infrared
observations to recalibrate secondary distance methods used in the
HST Key Project. These authors find
\begin{eqnarray*}
H_0=[74.3 \pm 1.5\text{(statistical)} \pm 2.1 \text{(systematic)}]
km s^{-1} Mpc^{-1}\\\nonumber\text{(Carnegie HP)},
\end{eqnarray*}
It can be observed through all plots (Figures \textbf{1-3}) for all
values of interacting parameter $d^2$ that $H(a)$ shows increasing
behavior which is consistent with the above observations.

\item In Figure \textbf{4} ($u=0.5$), it can be
observed that the EoS parameter starts from phantom region (with
comparatively large negative value) and goes towards lower negative
value of phantom region for all cases of interacting parameter. For
$u=-0.5$ (Figure \textbf{5}), it starts from quintessence phase  and
turns towards phantom region by crossing vacuum dominated era of the
universe for the cases ($d^2=0.02,~0.03$). However, it remains in
the phantom region for $d^2=0.04$. Also, Figure \textbf{6} provided
that EoS parameter starts comparatively high value of phantom region
and always remains in that region for all values of interacting
parameter. It can also be observed that EoS parameter attains high
phantom region with the increase of interacting parameter.

Moreover, Ade et al. \cite{R33} (Planck data) have put the following
constraints on the EoS parameter
\begin{eqnarray*}
\omega_{\Lambda}&=&-1.13^{+0.24}_{-0.25}~~~~~~~\text{(Planck+WP+BAO)},\\
\omega_{\Lambda}&=&-1.09\pm0.17,~~~\text{(Planck+WP+Union 2.1)}\\
\omega_{\Lambda}&=&-1.13^{+0.13}_{-0.14},~~~~~~\text{(Planck+WP+SNLS)},\\
\omega_{\Lambda}&=&-1.24^{+0.18}_{-0.19},~~~~~~\text{(Planck+WP+$H_0$)}.
\end{eqnarray*}
by implying different combination of observational schemes at $95\%$
confidence level. It can be seen from Figures \textbf{4-6} that the
EoS parameter also meets the above mentioned values for all cases of
interacting parameter which shows consistency of our results. The
above discussion shows that all the models provides fully support
the PDE phenomenon.

\item In Figure \textbf{7}, it is observed that GGPDE remains stable against small
perturbation at the present epoch as well as recent present epoch.
It can be viewed from Figure \textbf{8} ($u=-0.5$) that the GGPDE
model exhibits stability for all values of interacting parameter in
this scenario due to positive behavior of squared speed of sound. In
case of $u=1$ (Figure \textbf{9}), the squared speed of sound also
exhibits stability of the model for all cases of $d^2$.

\item The $\omega_{\Lambda}-\omega'_{\Lambda}$ plane for the current DE
model is constructed by plotting the $\omega'_{\Lambda}$ versus
$\omega_{\Lambda}$ for three different values of $u$ as shown in
Figures \textbf{10-12}. The specific values of other constant are
the same as above plots. Figures \textbf{10} and \textbf{12} provide
thawing region while Figure \textbf{11} exhibits freezing region.
The $\Lambda$CDM limit, i.e.,
$(\omega_{\Lambda},\omega'_{\Lambda})=(-1,0)$ only achieved for
$u=0.5$ with ($d^2=0$) as shown in Figure \textbf{10}. Also, Ade et
al. \cite{R33} have obtained the following constraints on
$w_{\Lambda}$ and $w'_{\Lambda}$:
\begin{eqnarray*}
\omega_{\Lambda}&=&-1.13^{+0.24}_{-0.25}~~~~~~~\text{(Planck+WP+BAO)},\\
\omega'_{\Lambda}&<&1.32,~~~~~~~~~~~~~~\text{(Planck+WP+BAO)}
\end{eqnarray*}
at $95\%$ confidence level. Also, other data with different
combinations of observational schemes such as (Planck+WP+Union 2.1)
and (Planck+WP+SNLS) favor the above constraints. In the present
case, the trajectories of $\omega'_{\Lambda}$ against
$\omega_{\Lambda}$ also meet the above mentioned values for all
cases of interacting parameter which shows consistency of our
results as shown in Figures \textbf{10-12}. Hence,
$\omega_{\Lambda}-\omega'_{\Lambda}$ plane provides consistent
behavior with the present day observations in all cases of $u$.

\item The $r-s$ plane corresponding to this scenario is shown in Figure
\textbf{13-15}. It is observed that the trajectories of $r-s$ plane
for all cases of interacting parameter corresponds to $\Lambda$CDM
model for $u=0.5$ as shown in Figure \textbf{13}. However, the
trajectories of $r-s$ meet $\Lambda$CDM limit for only $d^2=0$ (in
case of $u=-0.5$) and $d^2=0.02,~0.03$ (in case of $u=-0.5$) as
shown in Figures \textbf{14} and \textbf{15}, respectively. Also,
the trajectories coincide with the chaplygin gas model in all cases
of $u$.

\end{itemize}

\end{document}